\documentclass[sigchi]{acmart}

\usepackage{booktabs} 
\usepackage{subfiles}
\usepackage{balance}
\usepackage[utf8]{inputenc}

\copyrightyear{2019}
\acmYear{2019}
\setcopyright{acmlicensed}
\acmConference[CHI 2019]{CHI Conference on Human Factors in Computing Systems Proceedings}{May 4--9, 2019}{Glasgow, Scotland UK}
\acmBooktitle{CHI Conference on Human Factors in Computing Systems Proceedings (CHI 2019), May 4--9, 2019, Glasgow, Scotland UK}
\acmPrice{15.00}
\acmDOI{10.1145/3290605.3300247}
\acmISBN{978-1-4503-5970-2/19/05}

\begin{document}
\title[Moderation/Heterogeneity in Online Deliberation]{Effects of Moderation and Opinion Heterogeneity on Attitude towards the Online Deliberation Experience}

\author{Simon T. Perrault}
\authornote{Both authors contributed equally to the work.}
\orcid{0002-3105-9350}
\affiliation{%
  \institution{Yale-NUS College,}
}
\affiliation{%
  \institution{Singapore University of Technology and Design (SUTD)}
  \city{Singapore}
}
\affiliation{%
  \institution{Korea Advanced Institute of Science and Technology (KAIST)}
  \city{Korea}
}
\email{perrault.simon@gmail.com}

\author{Weiyu Zhang}
\authornote{Both authors contributed equally to the work.}
\affiliation{%
  \institution{Department of Communications and New Media}
  \institution{National University of Singapore (NUS)}
  \city{Singapore}
}
\email{weiyu.zhang@nus.edu.sg}

\renewcommand{\shortauthors}{Perrault and Zhang}

\begin{abstract}
Online deliberation offers a way for citizens to collectively discuss an issue and provide input for policy makers.
The overall experience of online deliberation can be affected by multiple factors.
We decided to investigate the effects of moderation and opinion heterogeneity on the perceived deliberation experience, by running the first online deliberation experiment in Singapore.
Our study took place in three months with three phases.
In phase 1, our 2,006 participants answered  a survey, that we used to create groups of different opinion heterogeneity.
During the second phase, 510 participants discussed about the population issue on the online platform we developed.
We gathered data on their online deliberation experience during phase 3.
We found out that higher levels of moderation negatively impact the experience of deliberation on perceived procedural fairness, validity claim and policy legitimacy; and that high opinion heterogeneity is important in order to get a fair assessment of the deliberation experience.
\end{abstract}

%
%




\maketitle
\section{Introduction}
Online deliberation is a young interdisciplinary field that studies deliberative processes with the use of information and communication technologies.
The deliberation process may last for days, or even weeks, and involves exchanging ideas on deliberation platforms.
The outcome of the discussion may be used to provide input for policy makers.

Compared to online discussion forums, online deliberation is more guided to have participants focus on specific and important topics, with the goal of achieving mutual understanding.
One of the key metrics used in deliberation is attitude change~\cite{Fishkin2011}, i.e. differences in the levels of agreement to specific topic statements before and after deliberation.
However, few studies~\cite{Chang2010} have examined participants' attitudes towards the deliberation experience itself, especially the legitimacy of this relatively new political practice.

Previous work in the field generally showed positive effects of online deliberation for democracy~\cite{Macintosh2004,Janssen2005} and was conducted in mature democracies, e.g. the USA~\cite{Fishkin2011}, or western European countries~\cite{Macintosh2004}.
Many factors may impact the deliberation outcomes, such as the topics of deliberation~\cite{Price2002}, the rules governing the deliberation~\cite{Zhang2014}, and the participants of deliberation activities~\cite{Zhang2015}.

In this paper, we present the first study conducted in Singapore, dubbed as an "authoritarian democracy"~\cite{Zhang2012}, in which we focus on the effect of moderation and opinion heterogeneity on attitudes towards the deliberation experience.

Our study that lasted for three months was done in three steps: (1) pre-deliberation survey, (2) deliberation itself and (3) post-deliberation survey.
Our pre-survey was done with 2,006 Singaporean citizens, representative of the country population.
For the deliberation phase, we developed our own open-source platform that enables simple discussions and provides educational material before each round of discussion.
A total of 510 participants followed phases 2 and 3.

Our results suggest that both moderation and opinion heterogeneity have an impact on attitudes towards the deliberation experience: (1) Participants who joined low moderation platforms perceived higher policy legitimacy, and thought of their discussion partners' claims as more valid, than those who joined high moderation platforms.
However, moderation made a marginally significant difference in perceived procedural fairness; (2) Participants who were put in the group with those mainly agreed with the policy gave higher scores in perceived procedural fairness than both the group with those who mainly disagreed with the policy and the group with those who were mixed in the views.
For validity claims and policy legitimacy, participants who were put in the group with those mainly disagreed with the policy gave lower scores than both the group with those who mainly agreed with the policy and the group with those who were mixed in their views.

To summarize, the contribution of this paper is three-fold:
\begin{itemize}
  \item We present an open-source deliberation platform that we developed,
  \item We describe the first online deliberation experiment in Singapore,
  \item We investigate the effect of moderation and opinion heterogeneity on attitudes towards the online deliberation experience.
\end{itemize}

\section{Related Work}
Our work specifically focuses on policy deliberation, which is different from other deliberative forum or more casual online discussions.
The goal of the whole experiment is also to provide citizens a space to discuss social policies, deliberate and provide input for policy makers.
We will first discuss the impact of moderation and opinion heterogeneity in online deliberation.
Then we will discuss how existing platforms dealt with both moderation and heterogeneity, and how our platform builds from these existing efforts.

\subsection{Moderation in Online Deliberation}
Moderation plays an important role in online discussion forums, and has been extensively discussed.
Some work, like Towne and Herbsleb~\cite{Towne2012}, advocate for a minimal amount of moderation to avoid trolling and flame wars.
Hanasono and Yang~\cite{Hanasono2016} pointed that even on a support group for racially discriminated people, an undisclosed proportion of messages were inappropriate or insulting.
Binns~\cite{Binns2012} described moderation as labour intensive and divided moderation to two different approaches.
Pre-moderation, means that comments need to be validated by a moderator to be published, which prevents battle between users and moderators.
With post-moderation, comments will be first published, then potentially moderated.
They also discussed the effect of adding a "karma" system on the platform, with a karma threshold required to pass pre-moderation.
Lazer et al.~\cite{Lazer2015} discussed moderators' strategies during their moderation work, while Epstein and Leshed~\cite{Epstein2016} highlighted the two main roles of moderators: (1) managing the stream of comments and (2) interacting with commenters~\cite{Edwards2002,Lopez2017}.

The impact of moderation on online discussion forums has been partially investigated.
For example, Meyer and Carey~\cite{Meyer2014} found that moderation would negatively affect the number of posts, as people are more likely to react to inappropriate posts, generating more activities and giving users a sense of virtual community.
Other negative effects include higher level of suspicion towards the deliberation process, self-censorship from users~\cite{Wright2009} and even potential exclusion of traditionally underrepresented populations~\cite{Trenel2009}, which would then hurt the representativeness of online discussion.

Wise et al.~\cite{Wise2006} conducted the only study so far which included moderation as an independent variable in their design, with two levels: moderation vs. no moderation.
However, their focus was on the moderation awareness, and how this awareness could lead to higher intent to participate in the process.

While these previous works guided decisions on how moderation should be done on our platform, e.g. to include underrepresented groups, none of them investigated the effect of moderation on participants' attitudes towards the deliberation experience.

\subsection{Opinion Heterogeneity in Online Deliberation}
Heterogeneity is a double-edge sword for online deliberation.
On one hand, it may be perceived as a threat to the harmony of social relationships, and would overall make it harder to create political mobilization~\cite{Mutz2002}.
On the other hand, Price et al.~\cite{Price2002} saw diverse opinions as benefiting democracy, because exposure to disagreement contributes to people's ability to generate reasons and improves argumentation.
Brundidge~\cite{Brundidge2010} proposed the inadvertency thesis to explain how people get exposed to heterogeneous opinions. Their findings suggest that users tend not to seek out political difference, but rather end up finding it inadvertently.
For example, Zhang and Chang~\cite{Zhang2014} reported that everyday political talk could be a good opportunity for people to find political difference.
Brundidge~\cite{Brundidge2010} also pointed out that the frequency of online political discussion is positively related to the heterogeneity of one's political discussion network.

There are multiple levels of heterogeneity, depending on the level of agreement or disagreement to attitude items.
High levels of disagreement or agreement would strongly affect participants' evaluations of deliberation processes~\cite{Stromer2009}.
This same work also suggests that one's satisfaction, potential reevaluation of opinion and expected future participation are not affected by low levels of disagreements, but high levels of disagreements could lower satisfaction and future engagement.
Finally, Wojcieszak and Price~\cite{Wojcieszak2012} conducted a study on perceived vs. actual disagreement. Their results show that perceived disagreement has strong effects on satisfaction or future participation, whereas objective disagreement does not.
Their results also confirm previous research on the negative effects of high disagreement on satisfaction~\cite{Zhang2015} and future engagement~\cite{Stromer2009}.

\subsection{Existing Deliberation Platforms}
HCI works on online deliberation usually focus on platforms and unique features implemented by each of these.
No prior work published in HCI venues has investigated the effect of moderation and opinion heterogeneity on the online deliberation experience.

Macintosh~\cite{Macintosh2004} defines three types of deliberation platforms: informative, consultative and participative.
Informative platforms are platforms used by governments to produce and deliver information for use by citizens.
Consultative platforms add a way for citizens to provide feedback on materials received.
Participative platforms allow citizens to actively engage in defining the process and content of policy-making.
Our work falls in this last category, which enables users to actively collaborate with each other, and discuss public issues. Our review does not cover aggregation platforms that aim at presenting different opinions and help users locate their own on the spectrum of opinions as well as reflect and refine their opinions, such as Opinion Space~\cite{Faridani2010}, ConsiderIt~\cite{Kriplean2011}, Reflect~\cite{Kriplean2012}, and Poli~\cite{Semaan2015}.

Our platform builds from existing collaborative platforms.
MIT Deliberatorium~\cite{Klein2011} takes users' input in the form of a deliberation map, a tree-structured network of posts representing a single issue, idea, or pro/con argument.
It is based on the Issue-Based Information System (IBIS) model of informal argumentation.
Arrangement of posts is based on topic, as opposed to time as seen on regular forums.
We used a similar hierarchical representation of discussion to make it easier to follow conversations.

Another early example of collaborative online deliberation was the Virtual Agora project~\cite{Muhlberger2005}.
Inspired by Deliberative Polls~\cite{Fishkin2005}, the system allowed participants to share web-based information and discuss using audio both in real-time and asynchronously.
The Virtual Agora~\cite{Muhlberger2005} project was able to enhance the salience of citizen identity, which in turn is correlated with higher political engagement.
PICOLA was a two-phase experiment regarding online citizen deliberation~\cite{Cavalier2009}.
Phase I of the study compared real-time online deliberation to face-to-face deliberation and the additional phase II explored the effects of asynchronous deliberations.
PICOLA came with an education phase for participants (implemented through an online reading room), a discussion phase (using a real-time online interaction), and a reflection phase (where participants could continue discussions through asynchronous online forum and surveys).
The results of the deliberation session of PICOLA were however not formally published.
Finally, Deme enabled collaborative drafting, focused discussion and decision making~\cite{Davies2004}.
Designed for small group interaction, it features meeting rooms where members could self-organize deliberative sessions.

Compared to aggregation platforms, our system offers an online space for citizens to discuss and deliberate.
We followed a similar process to PICOLA with an education phase, to mitigate potential lack of knowledgge of our participants on the discussion topics, and used a hierarchical structure of discussion roughly similar to MIT Deliberatorium~\cite{Klein2011}.

\subsection{Civic Tech in Asia}

Civic Tech has been rising in Asia during the last few years.
For example, Factful~\cite{Kim2015} is a platform that allows taxpayers to participate in the discussion of South Korea's government budget.
The platform allows users to read about budget planning and proposes news article and other sources.
This allows participants to understand the government choices, and be more critical while checking for facts.
Similarly, BudgetMap~\cite{Kim2016} is an issue-driven visualization platform that allows taxpayers to explore and tag the links between budget and social issues and was deployed in Seoul.
Finally, vTaiwan~\cite{Hsiao2018} is an open consultation process that brings citizens and government together in online and offline spaces, to deliberate and reach rough consensus on national issues, and to craft national digital legislation.
Our online deliberation experiment is one of the many efforts ongoing in the Asian Civic Tech scene.


\subsection{Hypotheses}
From the previous work, we hypothesize that both moderation and opinion heterogeneity will have significant impacts on deliberation outcomes, such as participants' perception of their deliberation experience.
However, due to the mixed findings from previous research, we are not able to make directional hypotheses.

\textbf{H1}: Different levels of moderation will lead to different levels of perceived procedural fairness, perceived validity claim, and perceived policy legitimacy.

\textbf{H2}: Different levels of opinion heterogeneity will lead to different levels of perceived procedural fairness, perceived validity claim, and perceived policy legitimacy.

\section{Process and Platform}
We developed our own platform which allowed us to run our experiment.
We will first discuss some of the challenges for running an online deliberation study and how the whole process was carried out, then we will present our platform and how the platform supports our process.
Finally, we will explain our participants sampling strategy.

\subsection{Challenges}
When we started the project, we faced two issues, highlighted by previous literature: (a) lack of knowledge of the issue discussed and (b) representativeness of the sample chosen for the study.
Lack of knowledge may lead to ill-informed opinions, which usually have low value for policy making~\cite{Fishkin2005}.
Lack of representativeness could undermine the potential acceptability of decisions made through deliberation to the whole society.

\paragraph{Lack of Knowledge.}
We addressed the lack of knowledge by running an early education phase.
During that phase, a team  of policy researchers, in consultation with the government agency in charge of policies related to population (our focus for this study).
The material produced explained each of the sub-issues, with numbers and sources, in a slideshow format, which was easy to read.
These slides focused on numbers, and we were careful about trying not to express official opinions, in order to avoid potential biases.
We pre-tested the sets of slides for each issue with small groups of participants.
The material was then inserted in the platform.
Participants had to review the material at least once and were free to review as many times as they wanted.

\paragraph{Representativeness.}
Another recurring challenge for online deliberation is the potential lack of representativeness of the participants pool~\cite{Elstub2014}.
Most of these projects sampled participants on a voluntary basis.
This leads to participants being potentially more engaged, and more tech-savy, excluding populations such as elderly or people with a low technology proficiency.

We addressed this issue by working with an online panel provider during the recruitment phase.
This provider had access to a panel of over 20,000 Singaporean citizens.
We set our demographic quotas to match the latest census data and monitored the sampling closely.
We manually corrected under-representation of given groups by sending e-mails to individuals of said groups.
As the deliberation phase lasted for three-weeks, we expected some drop-out.
Drop-out in itself is not an issue, unless only specific groups keep dropping out, by lack of time or resources.

To mitigate this issue, we provided incentives to sustain participation, especially for lower-income brackets.
Participants were thus compensated 50 SGD (37 USD) for taking part in the online deliberation phase.
They could also earn up to 30 additional SGD (22 USD) depending on their participation (for posting, reading, reacting).
While our original sample of 2,006 participants taking part in the pre-deliberation survey was overall representative of Singapore population on gender, ethnicity, housing, age, education and income criterion, we could not achieve a perfect representativeness for ou sample of 510 participants taking part in the deliberation and reflection phase.

\subsection{Deliberation Phases}\label{design-delib}
The study ran in three different phases: (1) recruitment/pre-deliberation phase, (2) deliberation phase, (3) reflection/post-deliberation phase.

\paragraph{Pre-deliberation Phase.}
During this phase, we recruited 2,006 participants (see Representativeness for more information).
Participants were invited to fill our pre-deliberation survey, which contained 40 close-ended questions about current policies regarding population.
At the end of the survey, participants were asked if they wanted to join the deliberation.
We sent an invitation to all participants from the respondents who were willing to join the deliberation.
This phase lasted for one month.

We used the answers from the pre-deliberation survey to group our participants into the three levels of opinion heterogeneity as explained in the Independent Variable subsection of the next section.

\paragraph{Deliberation Phase.}
During this three-weeks phase, participants were invited to join our platform.
Upon logging for the first time, participants were shown a video detailing the features of the platform and how to use it.
Note that participants were identified with a unique numerical ID on the platform, and no identifying information were published for privacy reasons.

Each week, citizens were invited to discuss one of the three sub-issues (fertility, foreign workforce, integration of new citizens).
Before proceeding to the discussion itself, participants would have to go through the education material for each sub-issue.
Discussion would start with the title, and a brief, carefully crafted opening message from the admins.
Administrators and moderators would not post any additional message throughout the process.
At the end of each discussion, an online poll was set up based on moderators' reading of opinions/suggestions that emerged during the discussion.
Participants were reminded to answer the poll before proceeding to the next sub-issue.

\paragraph{Post-deliberation Phase.}
During this one-month phase, participants were asked to fill a post-deliberation survey.
This survey essentially contained the same questions related to existing policies in Singapore, plus additional questions on their perception of their online deliberation experience.
Additionally, we asked some questions related to their experience with the platform.

We specifically used the post-deliberation survey to measure our three dependent variables: (1) perceived procedural fairness, (2) perceived validity claim and (3) perceived policy legitimacy.
The dependent variables are described in the subsection Dependent Variables of the next section.

\subsection{Platform}
\paragraph{Implementation.}
The platform was developed based on Vanilla Forums\footnote{https://vanillaforums.org/}, an open source forum platform.
The core of the platform is written in PHP, and data were stored in a MySQL database.
Since Vanilla Forums supports multiple plugins, we did tweak or develop new plugins to fit our needs.

\begin{figure}[!t]
  \includegraphics[width=.95\columnwidth]{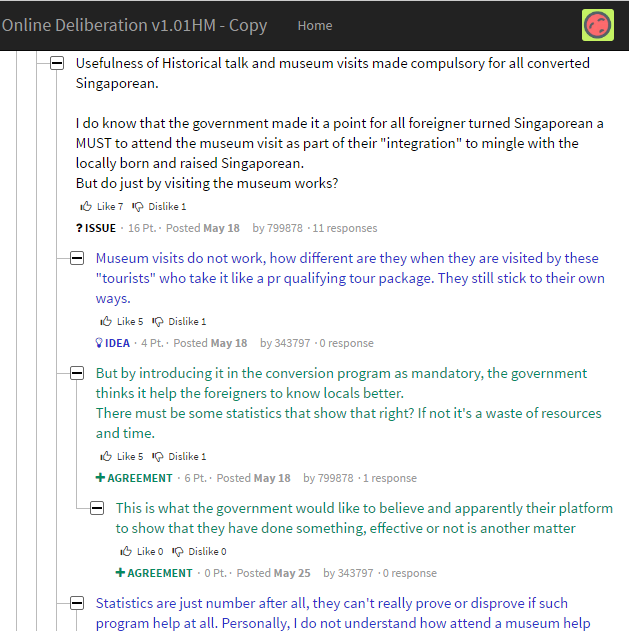}
  \caption{%
    Basic discussion interface. The discussion is structured as a tree, users can either expand or shrink nodes by clicking on the "+/-" button.
    \label{fig:discussion}%
  }
  \Description[Screenshort of the discussion platform.]{Screenshort of the discussion platform.}
\end{figure}

More advanced features were developed using JavaScript libraries.
We used Underscore.js, Vis.js and JQuery for the video display and graph creation/visualization tools.
The whole source code of our platform is available online at the following address: \url{http://onlinedeliberation.org/about}.
An example of a discussion can be seen on Figure~\ref{fig:discussion}.

\begin{figure}[!t]
  \includegraphics[width=.95\columnwidth]{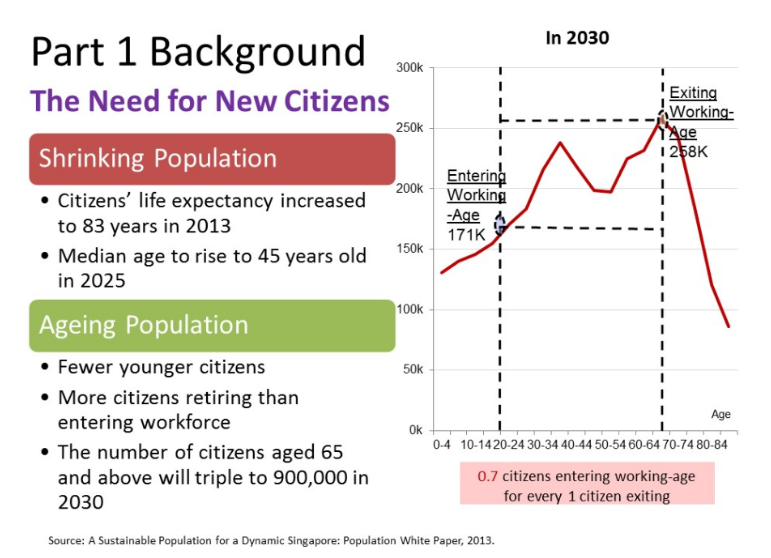}
  \caption{%
    Example of educational slide used for the sub-issue of Social Integration and New Citizens. This slide highlights the aging population issues, with less citizens entering the workforce and more exiting it by 2030. Each sub-issue came with a set of slides with data presented in a simple way.
    \label{fig:edu-slide}%
  }
  \Description[Educational Slide]{Example of a slide containing text explanations about aging population on the left, and a graph showing the evolution of the number of entering and exiting citizens in the workforce by 2030.}
\end{figure}

\paragraph{Educational Slides \& Quiz.}
Inspired by PICOLA~\cite{Cavalier2009}, the platform featured an educational phase at the beginning of each sub-issue discussion.
This phase consisted of a set of interactive PowerPoint slides (Figure~\ref{fig:edu-slide}) where each user would read through multiple educational slides and some self-paced interactive quiz that prompt citizen users to furnish their opinions about the issue.
The user was given the option of re-watching the slides and re-taking the quiz as many times as they wanted, to encourage deeper engagement with the learning material.

\begin{figure}[!b]
  \includegraphics[width=.95\columnwidth]{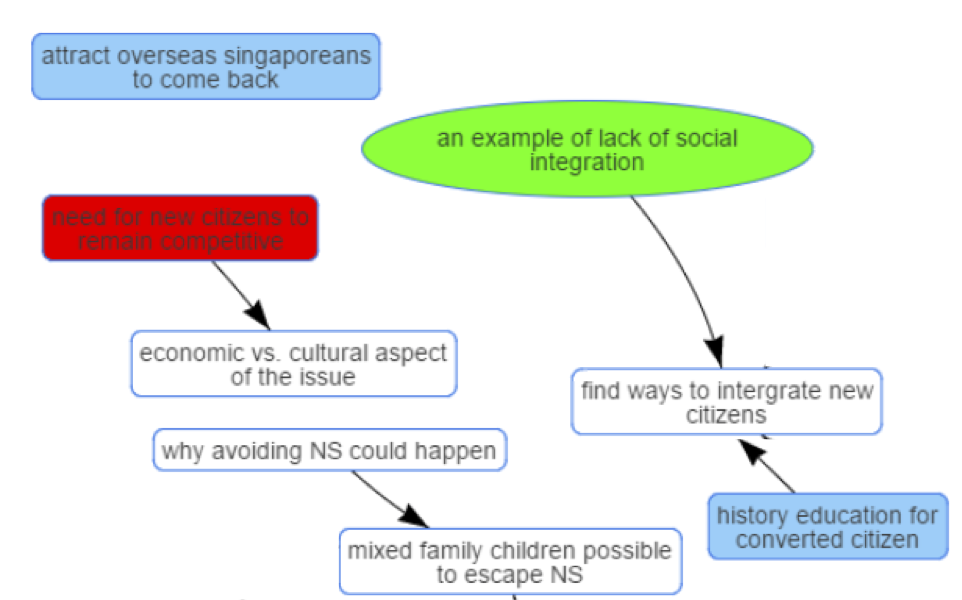}
  \caption{%
    Details of a deliberation graph generated to facilitate the discussion. Suggestions appear in boxes with a blue background, negative points in red and positive ones in green.
    \label{fig:graph-example}%
  }
  \Description[Deliberation Graph]{The deliberation graph is used to provide a visual representation of the discussion.}
\end{figure}

\paragraph{Deliberation Graph.}
Our platform also incorporated a deliberation graph (see Figure~\ref{fig:graph-example}), a mind map of salient posts (posted within a specific discussion) organized as an abstract topic model.
Inspired in part by the deliberation map from MIT deliberatorium~\cite{Klein2011}, the Deliberation Graph acts as a visual summary of an ongoing discussion.
But unlike the deliberatorium, the Deliberation Graph does not allow users to structure or populate the graph, and is instead operated by the moderator (for high moderation instances) who can use it to organize ongoing discussion along topic-hierarchies that they deem fit.
For low-moderation instances, the graph contains all the posts with their hierarchical relations and is automatically generated.
This flexibility with re-organization of arguments helps to further counteract the common knowledge problem, namely the tendency of groups to focus on the knowledge held by the majority and discount information held by the minority~\cite{Klein2011}.
The deliberation graph allowed for drawing shapes and directed connectors (each with changeable color and size attributes).
Once the moderator modifies the graph on a specific instance, changes are auto-saved and rendered instantly on the forum page for all users on the same instance.
Figure~\ref{fig:graph-example} shows an example of such a graph.
As a normal user, the graph allows zooming in and out on the graph and navigating from a node to a post that it refers to.
This mapping from node to post allows the moderator to frame posts in interesting and thought provoking ways without having to quote them verbatim.

\paragraph{IBIS based color coding of posts
.}
With the platform, we were also interested in providing argumentation support for users.
However, we did not want to make the experience of posting arguments counter-intuitive as is the case with MIT deliberatorium~\cite{Klein2011}.
We wanted to retain the normal time-wise sequential interface of traditional forums, but include interface element that could be used to organize posts based on the IBIS model of informal argumentation.
In terms of implementation, this meant the user was given a choice of categorizing each of their posts into either an issue, idea, agreement (pro argument) or disagreement (con argument).
A post belonged to each of these categories would have its own color code when displayed on the forum interface.
To further situate the IBIS structure, we used a nested style of indentation for posts, so that a reply to a post would appear at a level nested under the original post.
This way a user who has raised an issue or idea can have other users contribute pros and cons arguments under their post, offering the ability to have more focused discussions.

\paragraph{Gamification.}
We implemented both point systems and rewards in the platform.
As discussed in the design process section, on top of the baseline reimbursement, an additional discretionary reward was made available based on the level of user's participation during the deliberation.
The user can accrue participation points, by both submitting posts and by engaging with others' posts: 5 points gained by poster for every instance of their post getting 'liked', 5 points lost by poster for every instance of their post getting 'disliked', 5 points gained by Replying to posts, 2 points gained through liking/disliking another's post, 1 point gained for every instance of reading or viewing another's post.
The point system was tied to 3 milestone levels of participation, with each level unlocking additional monetary reward.
These milestones were capped at 1000, 2000 and 3000, with a commensurate reimbursement of 10SGD, 20SGD and 30SGD respectively.

We did not implement additional feature or content unlocking with these milestone levels, because this can lead to conflict with one of the core principles of deliberation namely accessible reasons~\cite{Gutmann1998}~(p.144).
This principle requires that the reasons and content used while deliberating should be accessible to all citizens.
Additionally, the use of social currency based incentives can lead to online hierarchy which impedes equal participation~\cite{Rosenberg2005}.
Keeping this in mind, we decided to not implement a leader board, although the individual participants' particulars were still made publically visible on the user's profile page.


\section{Online Deliberation Experiment}
In this experiment, we consider the effect of moderation (2 levels) and opinion heterogeneity (3 levels) on the perceived deliberation experience.
We rely on a between-subject design, and therefore split our participants into six different instances of our platform.

\subsection{Participants and Procedure}
We sent out an invitation to respondents from our pre-survey who consented to join the platform.
Prospective participants would receive an email containing a unique link to one of the six instances of the platform they were assigned to.
Among the 1200+ invitations we sent out, 510 participants logged into our platform at least once.
Among these 510 citizen users, 56\% were male; 86\% were ethnical Chinese; 55\% of people aged between 21 to 39 years old; 58\% had a university or high degree: and 57\% had a household monthly income below 10,000 SGD.
Compared to the general population, our participants tended to be male, Chinese, younger, better educated and having higher income.
A comparison between our participants and the singaporean population is shown in Table~\ref{tab:population}.

Following the pre-deliberation survey, during which participants provided their opinion on current policies, participants were assigned to one of our six instances (one per heterogeneity $\times$ moderation combination).
The deliberation phase was described in the Process and Platform section.
Each week, participants were invited to discuss one of the sub-issues, in that order:
\begin{enumerate}
    \item Singapore's Low Birthrate
    \item Foreign Domestic Workforce
    \item Integration of New Citizens
\end{enumerate}
After the deliberation, participants would answer a post-deliberation survey, in which we assessed the procedural fairness, validity claim and policy legitimacy as described in the Dependent Variables subsection.
\begin{table}[]
\caption{Population comparison between our participants and the singaporean population. Data computed and aggregated from \url{https://www.singstat.gov.sg/-/media/files/publications/reference/yearbook_2018/yos2018.pdf}.}
\label{tab:population}
\begin{tabular}{c|cc}
Criterion                                                             & Participants & SG Population \\ \hline
Male                                                                  & 56\%         & 49\%          \\
Chinese                                                               & 86\%         & 74.4\%        \\
Age 20-39                                                             & 55\%         & 28.4\%        \\
\begin{tabular}[c]{@{}c@{}}University\\ Degree\end{tabular}           & 58\%         & 30.7\%        \\
\begin{tabular}[c]{@{}c@{}}Income\\ \textless S\$ 10,000\end{tabular} & 57\%         & 63.4\%
\end{tabular}
\end{table}

\subsection{Independent Variables}
We considered two independent variables: moderation and opinion heterogeneity.

\paragraph{Moderation.}
We differentiated two levels of moderation, namely, \textbf{low} and \textbf{high}.
There are significant differences in terms of amount of human moderator's work.
In the \textbf{low} moderation conditions, moderators were instructed to keep the basic order of the discussion, such as posting the initial posts, removing duplicates and hiding offensive posts.
In the \textbf{high} moderation conditions, moderators have to do much more work: firstly, moderators greeted every participant with a welcome message that links to the help page which specifies discussion rules, when the participant posted the first post.
Secondly, moderators for the \textbf{high} moderation condition were in charge of generating and updating a deliberation graph that highlights good contributions and put them into a thought map at a daily frequency.
An automated deliberation map was also available for the \textbf{low} moderation conditions, which means all posts will automatically enter a network map without any moderator's interference.
Last but not least, moderators in the \textbf{high} moderation conditions were also managing inappropriate posts.
In fact, the only two instances in which we had to temporarily ban users both happened in high moderation conditions.

To summarize, the role of the moderator in the \textbf{high} moderation condition covers the two roles highlighted by Epstein and Leshd~\cite{Epstein2016} in addition to updating the deliberation graph daily.
In the \textbf{low} moderation condition, moderators would strictly stick to basic order keeping without directly interacting with the users.
Users were aware of the presence of the moderators by seeing moderators' posts such as the initial posts, and seeing messages such as "this post was hidden by a moderator" for hidden posts.

Four moderators were involved in moderating the discussions.
A 7-page moderation guide (available as supplementary material) was developed to provide standard instructions, including technical details, strategies to take when certain situations emerge, and templates to use for greeting, warning, banning users.
All moderators had to go through multiple rounds of training to ensure that they understand the moderation instructions fully.
All four moderators are communication scholars with at least a master degree in the major.
Three moderators who moderated the high moderation conditions rotated every week during the three-weeks discussion, to avoid conditions being confounded with moderators.
Moderators in the high moderation conditions spent at least two hours a day to complete the moderation tasks.
Thanks to the low workload, one moderator was tasked to moderate all low moderation conditions, to ensure the consistency among conditions.

\paragraph{Opinion Heterogeneity.}
To create our three groups of different opinion heterogeneity, we proceeded in two steps.
First, we generated a tripartite grouping based on the heterogeneity in group members' opinions towards policy positions, measured in the pre-deliberation survey.
The participants were first ranked on their level of agreement with policy positions (averaged across all three deliberation issues) and separated into 3 equal parts.
The first tertile of this partition was assigned a classification of Disagree (agreement score $\leq \frac{2.67}{5}$), second tertile a classification of Neutral (agreement score of $\frac{3}{5}$ or $\frac{3.33}{5}$) and the third quintile a classification of Agree (agreement score $ \geq \frac{3.67}{5}$).
\\
Second, we randomly selected the $\frac{2}{3}$ of the "Agree" users and $\frac{1}{3}$ of the "Neutral" users to create the Low Heterogeneity-Agree condition. We repeated this process for the Low Hete\-ro\-ge\-nei\-ty-Dis\-agree condition. The remaining users ($\frac{1}{3}$ of each original pool) was grouped as High Heterogeneity-Mixed.
Table~\ref{tab:heterogeneity} summarizes how the groups were created.
We will refer to each group according to their heterogeneity and opinion level.
As such, our three levels for heterogeneity are: \textbf{Low-Mainly Agree (L-A)}, \textbf{Low-Mainly Disagree (L-D)}, \textbf{High-Mixed (H-M)}.
\begin{table}[]
\caption{Group Composition for the Heterogeneity independent variable.}
\label{tab:heterogeneity}
\begin{tabular}{c|c}
Heterogeneity Level & Group Composition\\ \hline
Low-Agree              & $\frac{2}{3}$ agree, $\frac{1}{3}$ neutral \\
Low-Disagree           & $\frac{1}{3}$ neutral, $\frac{2}{3}$ disagree \\
High-Mixed              & $\frac{1}{3}$ agree, $\frac{1}{3}$ neutral, $\frac{1}{3}$ disagree
\end{tabular}
\end{table}

\subsection{Dependent Variables}\label{exp-dv}
We measured three dependent variables, all related to their overall experience of deliberation.

\paragraph{Perceived policy legitimacy}
This variable assessed participants' overall approval of the incorporation of deliberation in policymaking regarding population policies.
It was measured by four items: (a) I feel it is right for citizens to deliberate on taking action to control Singapore's population; (b) I feel citizen deliberation make just decisions to control Singapore's population; (c) I feel citizen deliberation can generate population policies that are worth of my support; and (d) I feel citizen deliberation can generate population policies that are likely to work.
A 5-point Likert scale was used (1 = strongly disagree, 5 = strongly agree) for all the questions.
Results of principal component analysis indicated that the four items loaded on one factor (Eigenvalue = 2.75, factor loadings between .78 and .89) and therefore, the four items were averaged to form one overall measure (Cronbach's $\alpha=.85$, $M=3.73$, $SD=.66$).

\paragraph{Perceived procedural fairness.}
We used a four-item measure in the post-deliberation survey to measure perceived procedural fairness. Building on prior studies on perceived speech conditions~\cite{Chang2010,Zhang2015,Zhang2014}, we asked participants how much they agree about these statements: (a) I had full capacity to freely raise questions about the population policies during the discussions; (b) I feel all of us had equal opportunities to express our opinions on the population issues during the discussions; (c) There was good balance in whose opinion about the population issues is being heard during the discussions; (d) I feel the people I discussed with gave a fair consideration to what I thought about the population issues. The same 5-point Likert scales were used. Results of principal component analysis indicated that the four items loaded on one factor (Eigenvalue = 2.90, factor loadings between .83 and .86) and therefore, the four items were averaged to form one overall measure (Cronbach's $\alpha=.87$, $M=2.31$, $SD=.99$).

\paragraph{Perceived validity claim.}
Four validity items were used to measure this concept~\cite{Chang2010}: (a) I think I understand fellow discussants' points about the population issues; (b) I think fellow discussants' discussions about the population issues were based on accurate facts; (c) I think fellow discussants communicated their arguments on the population issues in an appropriate way; and (d) I think fellow discussants expressed their sincere intentions to communicate with me on the population issues.
The same 5-point Likert scales were used.
All items loaded on one factor (Eigenvalue=2.71, factor loadings between .79 and .86) and hence they were combined into one variable (Cronbach's $\alpha=.84$, $M=3.68$, $SD=.64$).

\subsection{Data Analysis}
We ran a $2\times3$ between-subject study to empirically test the effects of \textit{Moderation} \{Low, High\} and \textit{Opinion Heterogeneity} \{ Low-Mainly Agree, Low-Mainly Disagree, High-Mixed\} on three dependent variables that measure participants' perceive experience with online deliberation.
We then performed a series of two-way ANCOVAs to analyze the effects of moderation and opinion heterogeneity.

\subsection{Covariates}
We ran a series of two-way ANOVAs with moderation and opinion heterogeneity as the two factors and with age, gender, ethnicity, education, housing type, and household income as dependent variables.
Only gender, ethnicity, and education showed some significant differences across heterogeneity conditions.
We decided to include these three variables as covariates in our main analyses.

\subsection{Online Activity Data}
We recorded a variety of online activies of our participants.
On average, each citizen user logged into our platform 11 times (range = 138; $SD=17$), liked 42 posts (range = 1087; $SD=125$), and disliked 2 posts (range = 28; $SD=15$).
Each of them posted an average of 12 posts (range = 305; $SD=36$) and viewed 84 posts (range = 1846; $SD=218$).
They clicked, on average, at least once the node in the deliberation graph (including both single clicks and double clicks; range = 41; $SD=3$).
Citizen users viewed our educational slides for an average of 1.43 times (range = 10; $SD=1$).
They accumulated an average of 590 points (range = 12505; $SD=1517$) through their platform activities.
Table~\ref{tab:descriptive} shows a summary of the posting activity by instances and conditions.
\begin{table}[!]
  \caption{Summary of the online deliberation phase in terms of Posts and Users. A user was considered active if they logged in at least once on the platform. L-A stands for Low Heterogeneity-Agree, H-M for High-Mixed and L-D fow Low-Disagree.}
  \label{tab:descriptive}
\begin{tabular}{cccc}
\textbf{Moderation} & \textbf{Heterogeneity} & \textbf{Posts} & \textbf{Active Users} \\ \hline
High       & L-A           & 453   & 74           \\
High       & H-M           & 893   & 90           \\
High       & L-D           & 1324  & 90           \\ \hline
Low        & L-A           & 529   & 71           \\
Low        & H-M           & 2162  & 94           \\
Low        & L-D           & 1147  & 91           \\ \hline
\textbf{Total}      & -            & 6508  & 510
\end{tabular}
\end{table}

\subsection{Results}
\begin{figure}[!]
  \includegraphics[width=.95\columnwidth]{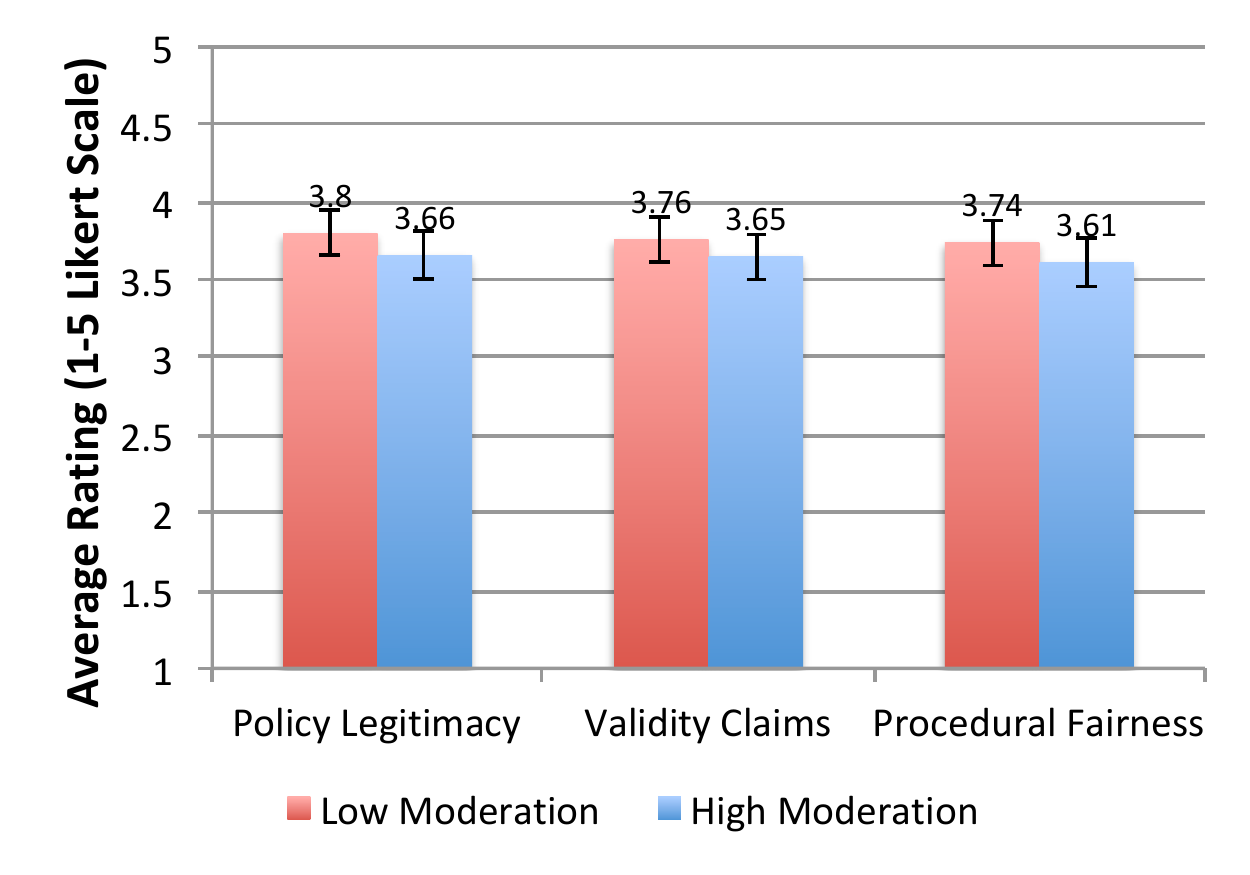}
  \caption{%
    Perceived policy legitimacy, validity claim and procedural fairness across moderation conditions. Error bars represent .95 confidence intervals.
    \label{fig:moderation}%
  }
  \Description[Results for Moderation]{High levels of moderation lead to lower scores for policy legitimacy, validity claim and procedural fairness.}
\end{figure}

\paragraph{Moderation.}
For the three dependent variables that gauge participants' perceptions about the online deliberation experience, both moderation and opinion heterogeneity made significant differences.
Moderation had significant impacts on perception of validity claim ($F_{1,437}=4.96$, $p<.05$, partial $\eta^{2}=.011$), perceived legitimacy of deliberation as a way of making population policies ($F_{1,437}=5.51$, $p<.05$, partial $\eta^{2}=.010$), and a marginally significant effect perception of procedural fairness ($F_{1,437}=3.08$, $p<.1$, partial $\eta^{2}=.007$).
Specifically, a consistent finding emerged: low moderation groups consistently had higher perception ratings than high moderation groups, as illustrated in Figure~\ref{fig:moderation}.
\textbf{H1 is supported}.

\begin{figure}[!]
  \includegraphics[width=.95\columnwidth]{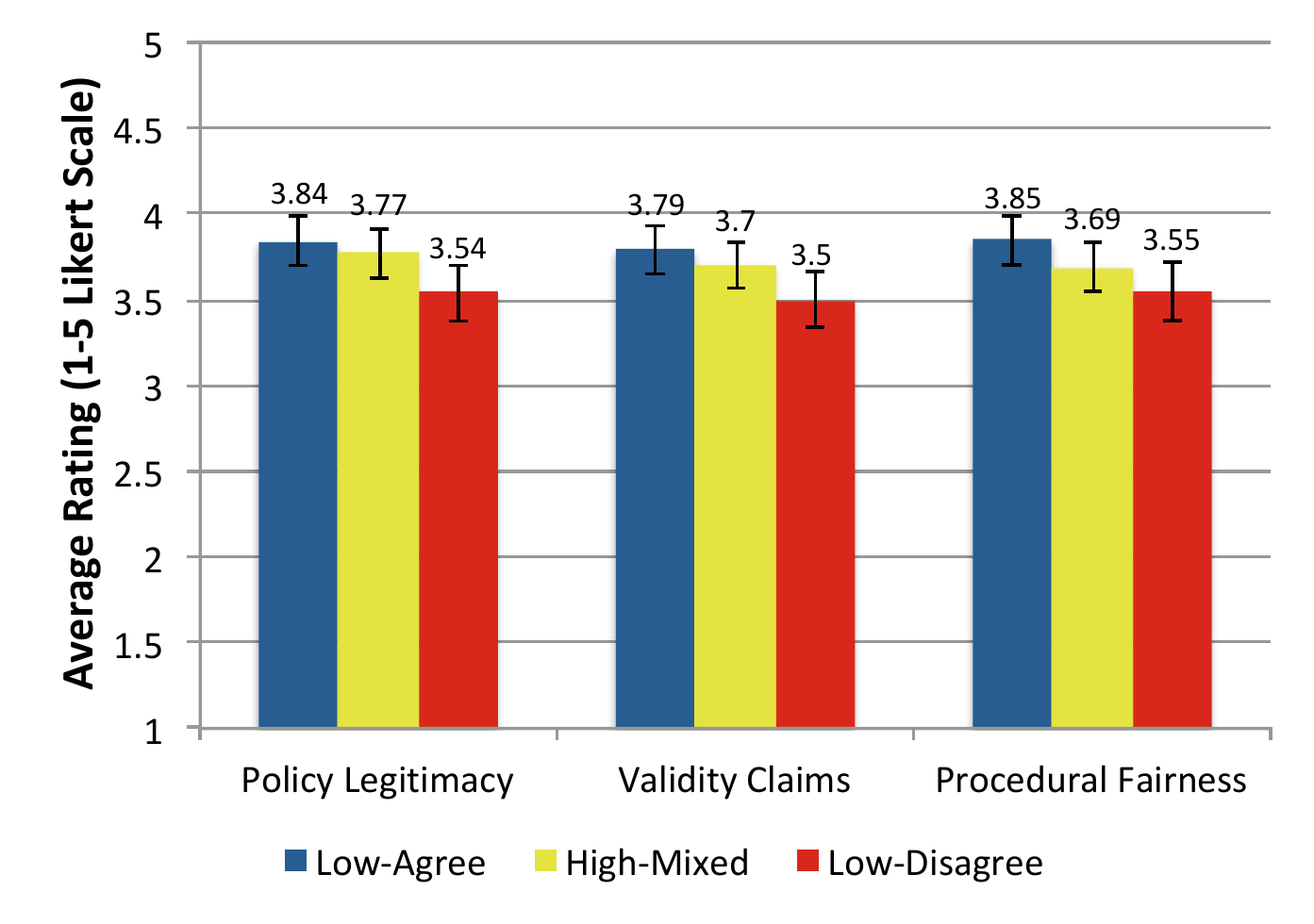}
  \caption{%
    Perceived policy legitimacy, validity claim and procedural fairness across opinion heterogeneity conditions. Error bars represent .95 confidence intervals.
    \label{fig:heterogeneity}%
  }
  \Description[Results for Opinion Heterogeneity]{The Low-Agree group show better perception in terms of policy legitimacy, validity claim and procedural fairness; followed by the High-Mixed group and Low-Disagree group.}
\end{figure}

\paragraph{Opinion Heterogeneity.}
Opinion heterogeneity also had significant impacts on perception of procedural fairness ($F_{2,437}=6.97$, $p<.01$, partial $\eta^{2}=.031$), perception of validity claim ($F_{2,437}=7.19$, $p<.01$, partial $\eta^{2}=.032$), and perception of policy legitimacy ($F_{2,437}=7.67$, $p<.01$, partial $\eta^{2}=.034$).
\\
Specifically, a consistent finding emerged with respect to perceptions of policy legitimacy and validity claim: the L-D group members perceived the lowest level of procedural fairness compared (policy legitimacy: $M=\frac{3.54}{5}$ and validity claim: $M=\frac{3.5}{5}$) to the L-A (policy legitimacy: $M=\frac{3.84}{5}, p<.001$; validity claim: $M =\frac{3.79}{5}, p<.001$) and H-M (policy legitimacy: $M=\frac{3.77}{5}, p<.01$; validity claim: $M=\frac{3.7}{5}, p<.01$) group members.
No significant differences were seen between the H-M and L-A group on these two measures.
\\
The pattern is slightly different in perceptions of procedural fairness: participants in the L-A group perceived the highest level of procedural fairness ($M=\frac{3.85}{5}$) compared to those in the H-M ($M=\frac{3.69}{5}, p<.05$) and L-D ($M=\frac{3.55}{5}, p<.001$) groups.
We did not observe any significant difference between the L-D and H-M groups on this measure.
The patterns can be seen in Figure~\ref{fig:heterogeneity}.
\textbf{H2 is supported}.

\paragraph{Interactions.}
We did not find any interaction effects between moderation and opinion heterogeneity (policy legitimacy: $p=.228$, validity claim: $p=.582$, procedural fairness: $p=.221$).

\section{Discussion and Guidelines}
We discuss the implications of our results for online deliberation, and provide guidelines to conduct future studies in the field.

\subsection{Challenges for Online Deliberation}
\paragraph{Use Educational Material.}
Regarding the lack of knowledge problem, we designed educational slides for both using the platform and each sub-issue.
Most participants did consult the slides more than once, and a majority of our participants (72\%) also found that "the educational slides were helpful".
Our participants' interest in educational slides suggest that future online deliberation works could benefit from applying a similar process.
It would also be interesting to quantify to which extent participants did learn from the slides, and how this might have impacted their opinions on issues.

\paragraph{Oversample Underrepresented Populations.}
Our second challenge was representativeness.
While we did start with a participant pool representative of Singapore's population in the pre-deliberation survey, we could not keep that representativeness.
We did still achieve a better representativeness compared to previous works based on voluntary participation.
Achieving representativeness in online deliberation requires a lot of work in view of the technological barriers to some groups of the population (e.g., the elderly).
To compensate for the attrition of less tech-savvy and older participants after the pre-deliberation survey, researchers may want to oversample participants of these underrepresented groups and match their numbers to the population quota after the initial survey.

\subsection{Keep Moderation to Low Levels}
Moderation had an impact on participants' perception of the deliberation experiment.
Lower levels of moderation leads to better perception of validity claim, perceived legitimacy of deliberation for policy making, and suggests that it may also affect procedural fairness.
Previous work also showed that lower levels of moderation could increase participation, in terms of number of posts~\cite{Meyer2014}, but higher levels of moderation lead to self-censorship~\cite{Wright2009} and exclude underrepresented populations from the deliberation process~\cite{Trenel2009}.
Given all these negative effects on discussion and perception of the deliberation process, we would argue to keep moderation to a strict minimum, in which moderators are simply required to prevent participants from resorting to insult or inappropriate behavior.

\subsection{Low Opinion Heterogeneity is Good?}
While statistical analyses suggests an effect of different heterogeneity levels on all three dependent variables, the trend that seems to emerge is that the instances of the platform where participants tended to agree (L-A) with current policies had a better perception of the deliberation process.
But does it mean that low opinion heterogeneity is good?
We would argue for the opposite: high opinion heterogeneity is not that bad.
If citizens already all agree/disagree to a policy, there is no need for deliberation.
Our findings show that at least H-M is better than L-D.
In fact, H-M showed no differences compared to L-A in two out of three outcomes, policy legitimacy and validity claim.
We would thus advocate to create groups of high heterogeneity with all kinds of opinions mixed, which should not only justify the need for deliberation but also reduce any evaluation of deliberation biased by one's agreement or disagreement to the policies.

\subsection{Implications for Online Discussion in General}
Our results are in line with past works on moderation (see above) which generally suggest to keep moderation to the minimum.
For instance, Towne and Herbsle~\cite{Towne2012} pointed out the importance of keeping at least some moderation to avoid trolling and flame wars.
In order to keep the basic order of discussions, moderation should also be quick.
But having a moderator working all the time on the platform is not always possible.
A solution is to use crowdsourced moderation~\cite{Lampe2014}, randomly assigning users the task to remove/hide offensive posts, which may act as a complement or replacement for centralized moderation.
However, this raises new challenges, such as small organized groups being able to moderate posts going against their ideas.

\subsection{Online Deliberation in Singapore}
This work is the first of its kind done in Singapore.
Because of the unique political context, the high moderation level could also trigger some mistrust or defiance towards the online deliberation process.
In a few isolated cases, some participants reacted very negatively to the presence of a moderator, but apart from these cases, we did not observe many negative reactions and/or signs or mistrust.

As shown by the results, the participants gave positive scores to all three of our dependent variables ($\frac{3.5}{5}$ and above).
In other questions of the post-deliberation survey, participants also reported "enjoying the discussions" (66.4\%) and found that the "discussions were interesting" (69.8\%).
The overall positive feedback on the experience suggests that there is space to conduct more of these studies in Singapore, giving citizens the opportunities to discuss online and provide input for local policy-makers.
As one of the participant reported: \textit{"I enjoyed very much my participation in this online deliberation and hope there will be more such activity/events organized. Perhaps, on other topics that concern Singapore as well, not just on Singapore's population issues"}.

\subsection{Limitations}
In our study, we used monetary incentives, as a way to address the representativeness issue~\cite{Elstub2014}.
Not providing incentives for a three-week long experiment would have meant excluding elderly, lower income and less tech-savvy people, as they tend to be the ones who are short of resources to support their participation in online deliberation.
Another possible effect of not providing financial incentives would be that people with strong opinions regarding the discussion topic would join voluntarily, while those with mild opinions would not care enough to participate.
This possible effect would render our experimental design futile, as we would not be able to attract enough participants with neutral opinions.
We acknowledge this limitation of our work, as voluntary participants are different from those who are motivated by extrinsic factors such as financial incentives.

\section{Conclusion}
We introduced our own open-source platform, which we built based on design recommendations from previous works.
During the deliberation phase, we evaluated how moderation and opinion heterogeneity may impact users' overall experience of deliberation.
We found that higher levels of moderation may degrade the overall experience.
We also found that opinion heterogeneity seems to be strongly linked to the perceptions of the participants, and therefore suggest to use high heterogeneity with mixed opinion in order to avoid any perception bias, positive or negative, based on one's existing opinions on the policies.
In terms of the Singapore context, our results suggest that despite of the unique political setting, participants were overall satisfied with their experience, and would be likely to support future deliberation practices like such.

\section{Acknowledgments}
This research is supported by Singapore's Ministry of Education (MOE2013-T2-1-105).
Views expressed are those of the authors alone and do not necessarily reflect opinions of the sponsoring agencies.
The authors want to thank Carol Soon, Leanne Chang, Shengdong Zhao, Tian Yang, Sanju Menon, Yang Wang, Jingjing Liang, Joseph Goh Seow Meng, Jeffrey Effendy, Atima Tharatipyakul and a group of undergraduate and graduate research assistants for their collaboration and assistance that made this research possible.

\bibliographystyle{ACM-Reference-Format}
\balance
\bibliography{delib}


\begin{thebibliography}{39}


\ifx \showCODEN    \undefined \def \showCODEN     #1{\unskip}     \fi
\ifx \showDOI      \undefined \def \showDOI       #1{#1}\fi
\ifx \showISBNx    \undefined \def \showISBNx     #1{\unskip}     \fi
\ifx \showISBNxiii \undefined \def \showISBNxiii  #1{\unskip}     \fi
\ifx \showISSN     \undefined \def \showISSN      #1{\unskip}     \fi
\ifx \showLCCN     \undefined \def \showLCCN      #1{\unskip}     \fi
\ifx \shownote     \undefined \def \shownote      #1{#1}          \fi
\ifx \showarticletitle \undefined \def \showarticletitle #1{#1}   \fi
\ifx \showURL      \undefined \def \showURL       {\relax}        \fi
\providecommand\bibfield[2]{#2}
\providecommand\bibinfo[2]{#2}
\providecommand\natexlab[1]{#1}
\providecommand\showeprint[2][]{arXiv:#2}

\bibitem[\protect\citeauthoryear{Binns}{Binns}{2012}]%
        {Binns2012}
\bibfield{author}{\bibinfo{person}{Amy Binns}.}
  \bibinfo{year}{2012}\natexlab{}.
\newblock \showarticletitle{DON'T FEED THE TROLLS! Managing troublemakers in
  magazines' online communities}.
\newblock \bibinfo{journal}{\emph{Journalism Practice}} \bibinfo{volume}{6},
  \bibinfo{number}{4} (\bibinfo{year}{2012}), \bibinfo{pages}{547--562}.
\newblock


\bibitem[\protect\citeauthoryear{Brundidge}{Brundidge}{2010}]%
        {Brundidge2010}
\bibfield{author}{\bibinfo{person}{Jennifer Brundidge}.}
  \bibinfo{year}{2010}\natexlab{}.
\newblock \showarticletitle{Encountering “difference” in the contemporary
  public sphere: The contribution of the Internet to the heterogeneity of
  political discussion networks}.
\newblock \bibinfo{journal}{\emph{Journal of Communication}}
  \bibinfo{volume}{60}, \bibinfo{number}{4} (\bibinfo{year}{2010}),
  \bibinfo{pages}{680--700}.
\newblock


\bibitem[\protect\citeauthoryear{Cavalier, Kim, and Zaiss}{Cavalier
  et~al\mbox{.}}{2009}]%
        {Cavalier2009}
\bibfield{author}{\bibinfo{person}{Robert Cavalier}, \bibinfo{person}{Miso
  Kim}, {and} \bibinfo{person}{Zachary~Sam Zaiss}.}
  \bibinfo{year}{2009}\natexlab{}.
\newblock \showarticletitle{Deliberative democracy, online discussion, and
  project PICOLA (Public informed citizen online assembly)}.
\newblock \bibinfo{journal}{\emph{Online deliberation: Design, research, and
  practice/Eds. Davies T., Gangadharan SP Stanford, CA: Center for the study of
  language and information}} (\bibinfo{year}{2009}), \bibinfo{pages}{71}.
\newblock


\bibitem[\protect\citeauthoryear{Chang and Jacobson}{Chang and
  Jacobson}{2010}]%
        {Chang2010}
\bibfield{author}{\bibinfo{person}{Leanne Chang} {and} \bibinfo{person}{Thomas
  Jacobson}.} \bibinfo{year}{2010}\natexlab{}.
\newblock \showarticletitle{Measuring Participation as Communicative Action: A
  Case Study of Citizen Involvement in and Assessment of a City's Smoking
  Cessation Policy-Making Process}.
\newblock \bibinfo{journal}{\emph{Journal of Communication}}
  \bibinfo{volume}{60}, \bibinfo{number}{4} (\bibinfo{year}{2010}),
  \bibinfo{pages}{660--679}.
\newblock
\urldef\tempurl%
\url{http://dx.doi.org/10.1111/j.1460-2466.2010.01508.x}
\showURL{%
\tempurl}


\bibitem[\protect\citeauthoryear{Davies, O'Connor, Cochran, Parker, Newman, and
  Tam}{Davies et~al\mbox{.}}{2009}]%
        {Davies2004}
\bibfield{author}{\bibinfo{person}{Todd Davies}, \bibinfo{person}{Brendan
  O'Connor}, \bibinfo{person}{Jonathan~J. Cochran, Alex Angiolilloand~Effrat},
  \bibinfo{person}{Andrew Parker}, \bibinfo{person}{Benjamin Newman}, {and}
  \bibinfo{person}{Aaron Tam}.} \bibinfo{year}{2009}\natexlab{}.
\newblock \showarticletitle{An online environment for democratic deliberation:
  motivations, principles, and design}.
\newblock  (\bibinfo{year}{2009}), \bibinfo{pages}{275--292}.
\newblock


\bibitem[\protect\citeauthoryear{Edwards}{Edwards}{2002}]%
        {Edwards2002}
\bibfield{author}{\bibinfo{person}{Arthur~R Edwards}.}
  \bibinfo{year}{2002}\natexlab{}.
\newblock \showarticletitle{The moderator as an emerging democratic
  intermediary: The role of the moderator in Internet discussions about public
  issues}.
\newblock \bibinfo{journal}{\emph{Information Polity}} \bibinfo{volume}{7},
  \bibinfo{number}{1} (\bibinfo{year}{2002}), \bibinfo{pages}{3--20}.
\newblock


\bibitem[\protect\citeauthoryear{Elstub and McLaverty}{Elstub and
  McLaverty}{2014}]%
        {Elstub2014}
\bibfield{author}{\bibinfo{person}{Stephen Elstub} {and} \bibinfo{person}{Peter
  McLaverty}.} \bibinfo{year}{2014}\natexlab{}.
\newblock \bibinfo{booktitle}{\emph{Deliberative democracy: Issues and cases}}.
\newblock \bibinfo{publisher}{Edinburgh University Press}.
\newblock


\bibitem[\protect\citeauthoryear{Epstein and Leshed}{Epstein and
  Leshed}{2016}]%
        {Epstein2016}
\bibfield{author}{\bibinfo{person}{Dmitry Epstein} {and} \bibinfo{person}{Gilly
  Leshed}.} \bibinfo{year}{2016}\natexlab{}.
\newblock \showarticletitle{The magic sauce: Practices of facilitation in
  online policy deliberation}.
\newblock \bibinfo{journal}{\emph{Journal of Public Deliberation}}
  \bibinfo{volume}{12}, \bibinfo{number}{1} (\bibinfo{year}{2016}),
  \bibinfo{pages}{4}.
\newblock


\bibitem[\protect\citeauthoryear{Faridani, Bitton, Ryokai, and
  Goldberg}{Faridani et~al\mbox{.}}{2010}]%
        {Faridani2010}
\bibfield{author}{\bibinfo{person}{Siamak Faridani}, \bibinfo{person}{Ephrat
  Bitton}, \bibinfo{person}{Kimiko Ryokai}, {and} \bibinfo{person}{Ken
  Goldberg}.} \bibinfo{year}{2010}\natexlab{}.
\newblock \showarticletitle{Opinion Space: A Scalable Tool for Browsing Online
  Comments}. In \bibinfo{booktitle}{\emph{Proceedings of the SIGCHI Conference
  on Human Factors in Computing Systems}} \emph{(\bibinfo{series}{CHI '10})}.
  \bibinfo{publisher}{ACM}, \bibinfo{address}{New York, NY, USA},
  \bibinfo{pages}{1175--1184}.
\newblock
\showISBNx{978-1-60558-929-9}
\urldef\tempurl%
\url{https://doi.org/10.1145/1753326.1753502}
\showDOI{\tempurl}


\bibitem[\protect\citeauthoryear{Fishkin}{Fishkin}{2011}]%
        {Fishkin2011}
\bibfield{author}{\bibinfo{person}{James~S Fishkin}.}
  \bibinfo{year}{2011}\natexlab{}.
\newblock \bibinfo{booktitle}{\emph{When the people speak: Deliberative
  democracy and public consultation}}.
\newblock \bibinfo{publisher}{Oxford University Press}.
\newblock


\bibitem[\protect\citeauthoryear{Fishkin and Luskin}{Fishkin and
  Luskin}{2005}]%
        {Fishkin2005}
\bibfield{author}{\bibinfo{person}{James~S Fishkin} {and}
  \bibinfo{person}{Robert~C Luskin}.} \bibinfo{year}{2005}\natexlab{}.
\newblock \showarticletitle{Experimenting with a democratic ideal: Deliberative
  polling and public opinion}.
\newblock \bibinfo{journal}{\emph{Acta politica}} \bibinfo{volume}{40},
  \bibinfo{number}{3} (\bibinfo{year}{2005}), \bibinfo{pages}{284--298}.
\newblock


\bibitem[\protect\citeauthoryear{Gutmann and Thompson}{Gutmann and
  Thompson}{1998}]%
        {Gutmann1998}
\bibfield{author}{\bibinfo{person}{Amy Gutmann} {and} \bibinfo{person}{Dennis~F
  Thompson}.} \bibinfo{year}{1998}\natexlab{}.
\newblock \bibinfo{booktitle}{\emph{Democracy and disagreement}}.
\newblock \bibinfo{publisher}{Harvard University Press}.
\newblock


\bibitem[\protect\citeauthoryear{Hanasono and Yang}{Hanasono and Yang}{2016}]%
        {Hanasono2016}
\bibfield{author}{\bibinfo{person}{Lisa~K Hanasono} {and} \bibinfo{person}{Fan
  Yang}.} \bibinfo{year}{2016}\natexlab{}.
\newblock \showarticletitle{Computer-mediated coping: Exploring the quality of
  supportive communication in an online discussion forum for individuals who
  are coping with racial discrimination}.
\newblock \bibinfo{journal}{\emph{Communication Quarterly}}
  \bibinfo{volume}{64}, \bibinfo{number}{4} (\bibinfo{year}{2016}),
  \bibinfo{pages}{369--389}.
\newblock


\bibitem[\protect\citeauthoryear{Hsiao, Lin, Tang, Narayanan, and Sarahe}{Hsiao
  et~al\mbox{.}}{2018}]%
        {Hsiao2018}
\bibfield{author}{\bibinfo{person}{Yu-Tang Hsiao}, \bibinfo{person}{Shu-Yang
  Lin}, \bibinfo{person}{Audrey Tang}, \bibinfo{person}{Darshana Narayanan},
  {and} \bibinfo{person}{Claudina Sarahe}.} \bibinfo{year}{2018}\natexlab{}.
\newblock \showarticletitle{vTaiwan: An Empirical Study of Open Consultation
  Process in Taiwan}.
\newblock  (\bibinfo{year}{2018}).
\newblock


\bibitem[\protect\citeauthoryear{Janssen and Kies}{Janssen and Kies}{2005}]%
        {Janssen2005}
\bibfield{author}{\bibinfo{person}{Davy Janssen} {and}
  \bibinfo{person}{Rapha{\"e}l Kies}.} \bibinfo{year}{2005}\natexlab{}.
\newblock \showarticletitle{Online forums and deliberative democracy}.
\newblock \bibinfo{journal}{\emph{Acta pol{\'\i}tica}} \bibinfo{volume}{40},
  \bibinfo{number}{3} (\bibinfo{year}{2005}), \bibinfo{pages}{317--335}.
\newblock


\bibitem[\protect\citeauthoryear{Kim, Ko, Jung, Lee, Kim, and Kim}{Kim
  et~al\mbox{.}}{2015}]%
        {Kim2015}
\bibfield{author}{\bibinfo{person}{Juho Kim}, \bibinfo{person}{Eun-Young Ko},
  \bibinfo{person}{Jonghyuk Jung}, \bibinfo{person}{Chang~Won Lee},
  \bibinfo{person}{Nam~Wook Kim}, {and} \bibinfo{person}{Jihee Kim}.}
  \bibinfo{year}{2015}\natexlab{}.
\newblock \showarticletitle{Factful: Engaging Taxpayers in the Public
  Discussion of a Government Budget}. In \bibinfo{booktitle}{\emph{Proceedings
  of the 33rd Annual ACM Conference on Human Factors in Computing Systems}}
  \emph{(\bibinfo{series}{CHI '15})}. \bibinfo{publisher}{ACM},
  \bibinfo{address}{New York, NY, USA}, \bibinfo{pages}{2843--2852}.
\newblock
\showISBNx{978-1-4503-3145-6}
\urldef\tempurl%
\url{https://doi.org/10.1145/2702123.2702352}
\showDOI{\tempurl}


\bibitem[\protect\citeauthoryear{Kim, Jung, Ko, Han, Lee, Kim, and Kim}{Kim
  et~al\mbox{.}}{2016}]%
        {Kim2016}
\bibfield{author}{\bibinfo{person}{Nam~Wook Kim}, \bibinfo{person}{Jonghyuk
  Jung}, \bibinfo{person}{Eun-Young Ko}, \bibinfo{person}{Songyi Han},
  \bibinfo{person}{Chang~Won Lee}, \bibinfo{person}{Juho Kim}, {and}
  \bibinfo{person}{Jihee Kim}.} \bibinfo{year}{2016}\natexlab{}.
\newblock \showarticletitle{BudgetMap: Engaging Taxpayers in the Issue-Driven
  Classification of a Government Budget}. In
  \bibinfo{booktitle}{\emph{Proceedings of the 19th ACM Conference on
  Computer-Supported Cooperative Work \& Social Computing}}
  \emph{(\bibinfo{series}{CSCW '16})}. \bibinfo{publisher}{ACM},
  \bibinfo{address}{New York, NY, USA}, \bibinfo{pages}{1028--1039}.
\newblock
\showISBNx{978-1-4503-3592-8}
\urldef\tempurl%
\url{https://doi.org/10.1145/2818048.2820004}
\showDOI{\tempurl}


\bibitem[\protect\citeauthoryear{Klein}{Klein}{2011}]%
        {Klein2011}
\bibfield{author}{\bibinfo{person}{M. Klein}.} \bibinfo{year}{2011}\natexlab{}.
\newblock \showarticletitle{The MIT deliberatorium: Enabling large-scale
  deliberation about complex systemic problems}. In
  \bibinfo{booktitle}{\emph{2011 International Conference on Collaboration
  Technologies and Systems (CTS)}}. \bibinfo{pages}{161--161}.
\newblock
\urldef\tempurl%
\url{https://doi.org/10.1109/CTS.2011.5928678}
\showDOI{\tempurl}


\bibitem[\protect\citeauthoryear{Kriplean, Morgan, Freelon, Borning, and
  Bennett}{Kriplean et~al\mbox{.}}{2011}]%
        {Kriplean2011}
\bibfield{author}{\bibinfo{person}{Travis Kriplean},
  \bibinfo{person}{Jonathan~T. Morgan}, \bibinfo{person}{Deen Freelon},
  \bibinfo{person}{Alan Borning}, {and} \bibinfo{person}{Lance Bennett}.}
  \bibinfo{year}{2011}\natexlab{}.
\newblock \showarticletitle{ConsiderIt: Improving Structured Public
  Deliberation}. In \bibinfo{booktitle}{\emph{CHI '11 Extended Abstracts on
  Human Factors in Computing Systems}} \emph{(\bibinfo{series}{CHI EA '11})}.
  \bibinfo{publisher}{ACM}, \bibinfo{address}{New York, NY, USA},
  \bibinfo{pages}{1831--1836}.
\newblock
\showISBNx{978-1-4503-0268-5}
\urldef\tempurl%
\url{https://doi.org/10.1145/1979742.1979869}
\showDOI{\tempurl}


\bibitem[\protect\citeauthoryear{Kriplean, Toomim, Morgan, Borning, and
  Ko}{Kriplean et~al\mbox{.}}{2012}]%
        {Kriplean2012}
\bibfield{author}{\bibinfo{person}{Travis Kriplean}, \bibinfo{person}{Michael
  Toomim}, \bibinfo{person}{Jonathan Morgan}, \bibinfo{person}{Alan Borning},
  {and} \bibinfo{person}{Andrew Ko}.} \bibinfo{year}{2012}\natexlab{}.
\newblock \showarticletitle{Is This What You Meant?: Promoting Listening on the
  Web with Reflect}. In \bibinfo{booktitle}{\emph{Proceedings of the SIGCHI
  Conference on Human Factors in Computing Systems}}
  \emph{(\bibinfo{series}{CHI '12})}. \bibinfo{publisher}{ACM},
  \bibinfo{address}{New York, NY, USA}, \bibinfo{pages}{1559--1568}.
\newblock
\showISBNx{978-1-4503-1015-4}
\urldef\tempurl%
\url{https://doi.org/10.1145/2207676.2208621}
\showDOI{\tempurl}


\bibitem[\protect\citeauthoryear{Lampe, Zube, Lee, Park, and Johnston}{Lampe
  et~al\mbox{.}}{2014}]%
        {Lampe2014}
\bibfield{author}{\bibinfo{person}{Cliff Lampe}, \bibinfo{person}{Paul Zube},
  \bibinfo{person}{Jusil Lee}, \bibinfo{person}{Chul~Hyun Park}, {and}
  \bibinfo{person}{Erik Johnston}.} \bibinfo{year}{2014}\natexlab{}.
\newblock \showarticletitle{Crowdsourcing civility: A natural experiment
  examining the effects of distributed moderation in online forums}.
\newblock \bibinfo{journal}{\emph{Government Information Quarterly}}
  \bibinfo{volume}{31}, \bibinfo{number}{2} (\bibinfo{year}{2014}),
  \bibinfo{pages}{317--326}.
\newblock


\bibitem[\protect\citeauthoryear{Lazer, Sokhey, Neblo, Esterling, and
  Kennedy}{Lazer et~al\mbox{.}}{2015}]%
        {Lazer2015}
\bibfield{author}{\bibinfo{person}{David~M Lazer}, \bibinfo{person}{Anand~E
  Sokhey}, \bibinfo{person}{Michael~A Neblo}, \bibinfo{person}{Kevin~M
  Esterling}, {and} \bibinfo{person}{Ryan Kennedy}.}
  \bibinfo{year}{2015}\natexlab{}.
\newblock \showarticletitle{Expanding the conversation: Multiplier effects from
  a deliberative field experiment}.
\newblock \bibinfo{journal}{\emph{Political Communication}}
  \bibinfo{volume}{32}, \bibinfo{number}{4} (\bibinfo{year}{2015}),
  \bibinfo{pages}{552--573}.
\newblock


\bibitem[\protect\citeauthoryear{L{\'o}pez~Garc{\'\i}a}{L{\'o}pez~Garc{\'\i}a}{2017}]%
        {Lopez2017}
\bibfield{author}{\bibinfo{person}{David L{\'o}pez~Garc{\'\i}a}.}
  \bibinfo{year}{2017}\natexlab{}.
\newblock \showarticletitle{Mediation Styles and Participants’ Perception of
  Success in Consultative Councils: The case of Guadalajara, Mexico}.
\newblock \bibinfo{journal}{\emph{Journal of Public Deliberation}}
  \bibinfo{volume}{13}, \bibinfo{number}{2} (\bibinfo{year}{2017}),
  \bibinfo{pages}{10}.
\newblock


\bibitem[\protect\citeauthoryear{Macintosh}{Macintosh}{2004}]%
        {Macintosh2004}
\bibfield{author}{\bibinfo{person}{Ann Macintosh}.}
  \bibinfo{year}{2004}\natexlab{}.
\newblock \showarticletitle{Characterizing e-participation in policy-making}.
  In \bibinfo{booktitle}{\emph{Proceedings of the 37th Annual Hawaii
  International Conference on System Sciences}}. IEEE, \bibinfo{pages}{10--pp}.
\newblock


\bibitem[\protect\citeauthoryear{Meyer and Carey}{Meyer and Carey}{2014}]%
        {Meyer2014}
\bibfield{author}{\bibinfo{person}{Hans~K Meyer} {and}
  \bibinfo{person}{Michael~Clay Carey}.} \bibinfo{year}{2014}\natexlab{}.
\newblock \showarticletitle{In Moderation: Examining how journalists' attitudes
  toward online comments affect the creation of community}.
\newblock \bibinfo{journal}{\emph{Journalism Practice}} \bibinfo{volume}{8},
  \bibinfo{number}{2} (\bibinfo{year}{2014}), \bibinfo{pages}{213--228}.
\newblock


\bibitem[\protect\citeauthoryear{Muhlberger}{Muhlberger}{2005}]%
        {Muhlberger2005}
\bibfield{author}{\bibinfo{person}{Peter Muhlberger}.}
  \bibinfo{year}{2005}\natexlab{}.
\newblock \showarticletitle{The Virtual Agora Project: A research design for
  studying democratic deliberation}.
\newblock \bibinfo{journal}{\emph{Journal of Public Deliberation}}
  \bibinfo{volume}{1}, \bibinfo{number}{1} (\bibinfo{year}{2005}),
  \bibinfo{pages}{5}.
\newblock


\bibitem[\protect\citeauthoryear{Mutz}{Mutz}{2002}]%
        {Mutz2002}
\bibfield{author}{\bibinfo{person}{Diana~C Mutz}.}
  \bibinfo{year}{2002}\natexlab{}.
\newblock \showarticletitle{The consequences of cross-cutting networks for
  political participation}.
\newblock \bibinfo{journal}{\emph{American Journal of Political Science}}
  (\bibinfo{year}{2002}), \bibinfo{pages}{838--855}.
\newblock


\bibitem[\protect\citeauthoryear{Price, Cappella, and Nir}{Price
  et~al\mbox{.}}{2002}]%
        {Price2002}
\bibfield{author}{\bibinfo{person}{Vincent Price}, \bibinfo{person}{Joseph~N
  Cappella}, {and} \bibinfo{person}{Lilach Nir}.}
  \bibinfo{year}{2002}\natexlab{}.
\newblock \showarticletitle{Does disagreement contribute to more deliberative
  opinion?}
\newblock \bibinfo{journal}{\emph{Political communication}}
  \bibinfo{volume}{19}, \bibinfo{number}{1} (\bibinfo{year}{2002}),
  \bibinfo{pages}{95--112}.
\newblock


\bibitem[\protect\citeauthoryear{Rosenberg}{Rosenberg}{2005}]%
        {Rosenberg2005}
\bibfield{author}{\bibinfo{person}{Shawn Rosenberg}.}
  \bibinfo{year}{2005}\natexlab{}.
\newblock \showarticletitle{The Empirical Study of Deliberative Democracy:
  Setting a Research Agenda}.
\newblock \bibinfo{journal}{\emph{Acta Politica}} \bibinfo{volume}{40},
  \bibinfo{number}{2} (\bibinfo{date}{01 Jul} \bibinfo{year}{2005}),
  \bibinfo{pages}{212--224}.
\newblock
\showISSN{1741-1416}
\urldef\tempurl%
\url{https://doi.org/10.1057/palgrave.ap.5500105}
\showDOI{\tempurl}


\bibitem[\protect\citeauthoryear{Semaan, Faucett, Robertson, Maruyama, and
  Douglas}{Semaan et~al\mbox{.}}{2015}]%
        {Semaan2015}
\bibfield{author}{\bibinfo{person}{Bryan Semaan}, \bibinfo{person}{Heather
  Faucett}, \bibinfo{person}{Scott~P. Robertson}, \bibinfo{person}{Misa
  Maruyama}, {and} \bibinfo{person}{Sara Douglas}.}
  \bibinfo{year}{2015}\natexlab{}.
\newblock \showarticletitle{Designing Political Deliberation Environments to
  Support Interactions in the Public Sphere}. In
  \bibinfo{booktitle}{\emph{Proceedings of the 33rd Annual ACM Conference on
  Human Factors in Computing Systems}} \emph{(\bibinfo{series}{CHI '15})}.
  \bibinfo{publisher}{ACM}, \bibinfo{address}{New York, NY, USA},
  \bibinfo{pages}{3167--3176}.
\newblock
\showISBNx{978-1-4503-3145-6}
\urldef\tempurl%
\url{https://doi.org/10.1145/2702123.2702403}
\showDOI{\tempurl}


\bibitem[\protect\citeauthoryear{Stromer-Galley and Muhlberger}{Stromer-Galley
  and Muhlberger}{2009}]%
        {Stromer2009}
\bibfield{author}{\bibinfo{person}{Jennifer Stromer-Galley} {and}
  \bibinfo{person}{Peter Muhlberger}.} \bibinfo{year}{2009}\natexlab{}.
\newblock \showarticletitle{Agreement and disagreement in group deliberation:
  Effects on deliberation satisfaction, future engagement, and decision
  legitimacy}.
\newblock \bibinfo{journal}{\emph{Political communication}}
  \bibinfo{volume}{26}, \bibinfo{number}{2} (\bibinfo{year}{2009}),
  \bibinfo{pages}{173--192}.
\newblock


\bibitem[\protect\citeauthoryear{Towne and Herbsleb}{Towne and
  Herbsleb}{2012}]%
        {Towne2012}
\bibfield{author}{\bibinfo{person}{W~Ben Towne} {and} \bibinfo{person}{James~D
  Herbsleb}.} \bibinfo{year}{2012}\natexlab{}.
\newblock \showarticletitle{Design considerations for online deliberation
  systems}.
\newblock \bibinfo{journal}{\emph{Journal of Information Technology \&
  Politics}} \bibinfo{volume}{9}, \bibinfo{number}{1} (\bibinfo{year}{2012}),
  \bibinfo{pages}{97--115}.
\newblock


\bibitem[\protect\citeauthoryear{Tr{\'e}nel}{Tr{\'e}nel}{2009}]%
        {Trenel2009}
\bibfield{author}{\bibinfo{person}{Matthias Tr{\'e}nel}.}
  \bibinfo{year}{2009}\natexlab{}.
\newblock \showarticletitle{Facilitation and inclusive deliberation}.
\newblock \bibinfo{journal}{\emph{Online deliberation: Design, research, and
  practice}} (\bibinfo{year}{2009}), \bibinfo{pages}{253--257}.
\newblock


\bibitem[\protect\citeauthoryear{Wise, Hamman, and Thorson}{Wise
  et~al\mbox{.}}{2006}]%
        {Wise2006}
\bibfield{author}{\bibinfo{person}{Kevin Wise}, \bibinfo{person}{Brian Hamman},
  {and} \bibinfo{person}{Kjerstin Thorson}.} \bibinfo{year}{2006}\natexlab{}.
\newblock \showarticletitle{Moderation, response rate, and message
  interactivity: Features of online communities and their effects on intent to
  participate}.
\newblock \bibinfo{journal}{\emph{Journal of Computer-Mediated Communication}}
  \bibinfo{volume}{12}, \bibinfo{number}{1} (\bibinfo{year}{2006}),
  \bibinfo{pages}{24--41}.
\newblock


\bibitem[\protect\citeauthoryear{Wojcieszak and Price}{Wojcieszak and
  Price}{2012}]%
        {Wojcieszak2012}
\bibfield{author}{\bibinfo{person}{Magdalena~E Wojcieszak} {and}
  \bibinfo{person}{Vincent Price}.} \bibinfo{year}{2012}\natexlab{}.
\newblock \showarticletitle{Perceived versus actual disagreement: Which
  influences deliberative experiences?}
\newblock \bibinfo{journal}{\emph{Journal of communication}}
  \bibinfo{volume}{62}, \bibinfo{number}{3} (\bibinfo{year}{2012}),
  \bibinfo{pages}{418--436}.
\newblock


\bibitem[\protect\citeauthoryear{Wright}{Wright}{2009}]%
        {Wright2009}
\bibfield{author}{\bibinfo{person}{Scott Wright}.}
  \bibinfo{year}{2009}\natexlab{}.
\newblock \showarticletitle{The role of the moderator: Problems and
  possibilities for government-run online discussion forums}.
\newblock \bibinfo{journal}{\emph{Online deliberation: Design, research, and
  practice}} (\bibinfo{year}{2009}), \bibinfo{pages}{233--242}.
\newblock


\bibitem[\protect\citeauthoryear{Zhang}{Zhang}{2012}]%
        {Zhang2012}
\bibfield{author}{\bibinfo{person}{Weiyu Zhang}.}
  \bibinfo{year}{2012}\natexlab{}.
\newblock \showarticletitle{The effects of political news use, political
  discussion and authoritarian orientation on political participation:
  Evidences from Singapore and Taiwan}.
\newblock \bibinfo{journal}{\emph{Asian Journal of Communication}}
  \bibinfo{volume}{22}, \bibinfo{number}{5} (\bibinfo{year}{2012}),
  \bibinfo{pages}{474--492}.
\newblock


\bibitem[\protect\citeauthoryear{Zhang}{Zhang}{2015}]%
        {Zhang2015}
\bibfield{author}{\bibinfo{person}{Weiyu Zhang}.}
  \bibinfo{year}{2015}\natexlab{}.
\newblock \showarticletitle{Perceived procedural fairness in deliberation:
  Predictors and effects}.
\newblock \bibinfo{journal}{\emph{Communication Research}}
  \bibinfo{volume}{42}, \bibinfo{number}{3} (\bibinfo{year}{2015}),
  \bibinfo{pages}{345--364}.
\newblock


\bibitem[\protect\citeauthoryear{Zhang and Chang}{Zhang and Chang}{2014}]%
        {Zhang2014}
\bibfield{author}{\bibinfo{person}{Weiyu Zhang} {and} \bibinfo{person}{Leanne
  Chang}.} \bibinfo{year}{2014}\natexlab{}.
\newblock \showarticletitle{Perceived speech conditions and disagreement of
  everyday talk: A proceduralist perspective of citizen deliberation}.
\newblock \bibinfo{journal}{\emph{Communication Theory}} \bibinfo{volume}{24},
  \bibinfo{number}{2} (\bibinfo{year}{2014}), \bibinfo{pages}{124--145}.
\newblock


\end{thebibliography}

\end{document}